\documentclass[showpacs,preprintnumbers,amsmath,amssymb,superscriptaddress]{revtex4}

\usepackage{amsfonts}
\usepackage{graphicx}
\usepackage{dcolumn}
\usepackage{bm}

\begin{document}


\title{Blume-Emery-Griffiths dynamics in social networks}

\author{Yao-Hui Yang}
\affiliation{Department of Mathematics and Physics, Chongqing University of Science and Technology, Chongqing $401331$, China}

\date{\today}

\begin{abstract}
We introduce the Blume-Emery-Griffiths (BEG) model in a social networks to describe the three-state dynamics of opinion formation. It shows that the probability distribution function of the time series of opinion is a Gaussian-like distribution. We also study the response of BEG model to the external periodic perturbation. One can observe that both the interior thermo-noise and the external field result in phase transition, which is a split phenomena of the opinion distributions. It is opposite between the effect acted on the opinion systems of the amplitude of the external field and of the thermo-noise.
\end{abstract}


\maketitle

\section{\label{sec1}INTRODUCTION}

Over the last few years, the study of opinion formation in complex
networks has attracted a growing amount of works and becomes the
major trend of sociophysics~\cite{Intro-1}. Many models have been
proposed, like those of Deffuant~\cite{Intro-2},
Galam~\cite{Intro-3}, Krause-Hegselmann (KH)~\cite{Intro-4}, and
Sznajd~\cite{Intro-5}. But most models in the literature consider
two-state opinion agents, in favor ($+1$) or against ($-1$) about a
certain topic. In the Galam's majority rule and the Sznajd's
updating rule, the interaction between the agents is randomly
changed during the evolution, and the time to reach consensus is
associated with the initial traction $p$ of $+1$ state. The
consensus time $T$ reaches its maximal value at $p=0.5$. In the
Sznajd model, a pair of nearest neighbors convinces its neighbors to
adopt the pair opinion if and only if both members have the same
opinion. Otherwise the pair and its neighbors do not change opinion.
In the KH consensus model, the opinions between $0$ and $1$ and a
confidence bound parameter is introduced. The agent $i$ would take
the average opinion of all neighboring agents that are within a
confidence bound during the evolution. In the Deffuant model, the
opinion of two randomly selected neighboring agents $i$ and $j$
would remain unchanged, if their opinions $\sigma_{i}$ and
$\sigma_{j}$ differ by more than a fixed threshold parameter.
Otherwise, each opinion moves into the direction of the other by an
amount $\mu\times\mid\sigma_{i}-\sigma_{j}\mid$.

Additionally, complex networks have received much attention in recent years. Topologically, a network is consisted of nodes and links. The complex network models, such as the lattice network, the random network~\cite{Intr_6,Intr_7,Intr_8}, the small-world network~\cite{Intr_9,Intr_10}, and the scale-free network~\cite{Intr_11}, are studied in many branches of science. It is meaningful to mention that opinion formation models are set up in complex networks.

In the present work, we investigate the implication of a social
network in a stochastic opinion formation model. We first introduce
the Blume-Emery-Griffiths (BEG) model~\cite{Intr_12,Intr_13,Intr_14}
to describe the dynamics of opinion formation, and the model of
complex networks we used is social network which is more reality.
Our simulation focuses on the average opinion for different
situation. And we also simulated the system under the influence of
external field.

In the rest of this paper we will give a description of this dynamic model and how to generate the underlying networks. In Sec.\ref{sec3}, we show the simulation results without external filed. In Sec.~\ref{sec4} we present the results with the influence of external field. The final section presents further discussion and conclusion.

\section{The model\label{sec2}}

Generally speaking, social networks include some essential characteristics, such as short average path lengths, high clustering, assortative mixing~\cite{Model-1,Model-2}, the existence of community structure, and broad degree distributions~\cite{Model-3,Model-4}. As a result, we use Riitta Toivonen's social network model in our present work~\cite{Model-5}. This network is structured by two processes: $1)$ attachment to random vertices, and $2)$ attachment to the neighborhood of the random vertices, giving rise to implicit preferential attachment. These processes give rise to essential characteristics for social networks. The second process gives rise to assortativity, high clustering and community structure. The degree distribution is also determined by the number of edges generated by the second process for each random attachment.

\begin{figure}
\includegraphics[width=0.5\textwidth]{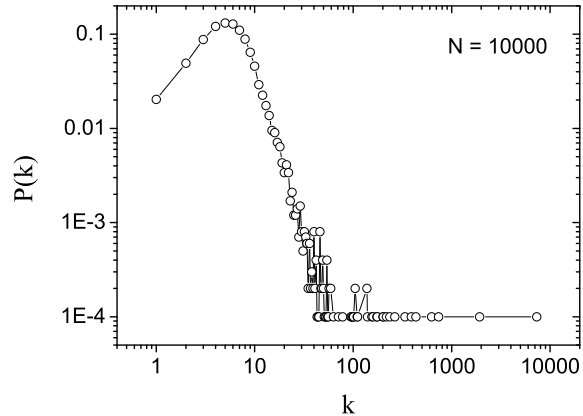}
\caption{Degree distribution of networks with $N=10000$. Result is averages over $20$ simulation runs. The number of initial contacts is distributed as $p(n_{init}=1)=0.25$, $p(n_{init}=2)=0.75$, and the number of secondary contacts
from each initial contact $n_{2nd}\sim U[0, 3]$. }
\label{degree}
\end{figure}

In this paper, the network is grown from a chain with $10$ nodes. The number of initial contacts is distributed as $p(n_{init}=1)=0.25$, $p(n_{init}=2)=0.75$, and the number of secondary contacts from each initial contact $n_{2nd}\sim U[0, 3]$ (uniformly distributed between $0$ and $3$). The total number of nodes in the social network structure is $N=10000$. The degree distribution of simulated networks is displayed in Fig.~\ref{degree}. We note that the degree distributon $P(k)$ is a power-law functional form and a peak around the degree $k=5$, also that consistent with real world observations~\cite{Intr_11,Model-6}.

Now, we consider a system with $N$ agents, which is represented by nodes on a social network. For each node, we consider three states which are represented by $+1$, $0$, and $-1$. A practical example could be the decision to agree $\sigma_{i}(t)=+1$, disagree $\sigma_{i}(t)=-1$, or neutral $\sigma_{i}(t)=0$. The states are updated according to the stochastic parallel spin-flip dynamics defined by the transition probabilities
\begin{equation}
Prob \left( \sigma_{i,t+1} = s'|\sigma_{N}(t) \right) = \frac {\exp \left\{ -\beta \epsilon_{i} \left[ s'|\sigma_{N}(t) \right] \right\}} {\sum_{s}\exp
\left\{-\beta\epsilon_{i}[s|\sigma_{N}(t)]\right\}}
\label{eq01}
\end{equation}
where $s, s' \in \{+1, 0, -1\}$, and $\beta=a/T$, $a$ represents the active degree of system, defined as $a = \left< \sigma_{N}^{2}(t) \right>$. The energy potential $\epsilon_{i} \left[ s|\sigma_{N}(t) \right]$ is
defined by
\begin{equation}
\epsilon_{i} \left[ s|\sigma_{N}(t) \right] = - s h_{i} \left( \sigma_{N}(t) \right) - s^{2}\theta_{i} \left( \sigma_{N}(t) \right),
\label{eq02}
\end{equation}
where the following local field in node $i$ carries all information
\begin{eqnarray}
h_{N,i}(t)&=&\sum_{j\neq i}J_{ij}\sigma_{j}(t), \nonumber\\
\theta_{N,i}(t)&=&\sum_{j\neq i}K_{ij}\sigma_{j}^{2}(t). \nonumber
\end{eqnarray}
Here, we define coupling $J_{ij}$ and $K_{ij}$ are positive numbers less than or equal to $1$, and with Gaussian distribution. $h_{N,i}(t)$ represents the time dependent interaction strengths between the node $i$ and his $n_{i}$ nearest neighboring nodes. $\theta_{N,i}(t)$ instead the strengths of feedback and $T$ is interior thermo-noise. So the average opinion is defined by
\begin{equation}
r(t) = \frac{1}{N} \sum_{j=1}^{N} \sigma_{j}(t).
\label{eq03}
\end{equation}

\section{Simulation results\label{sec3}}

\begin{figure}
\includegraphics[width=0.4\textwidth]{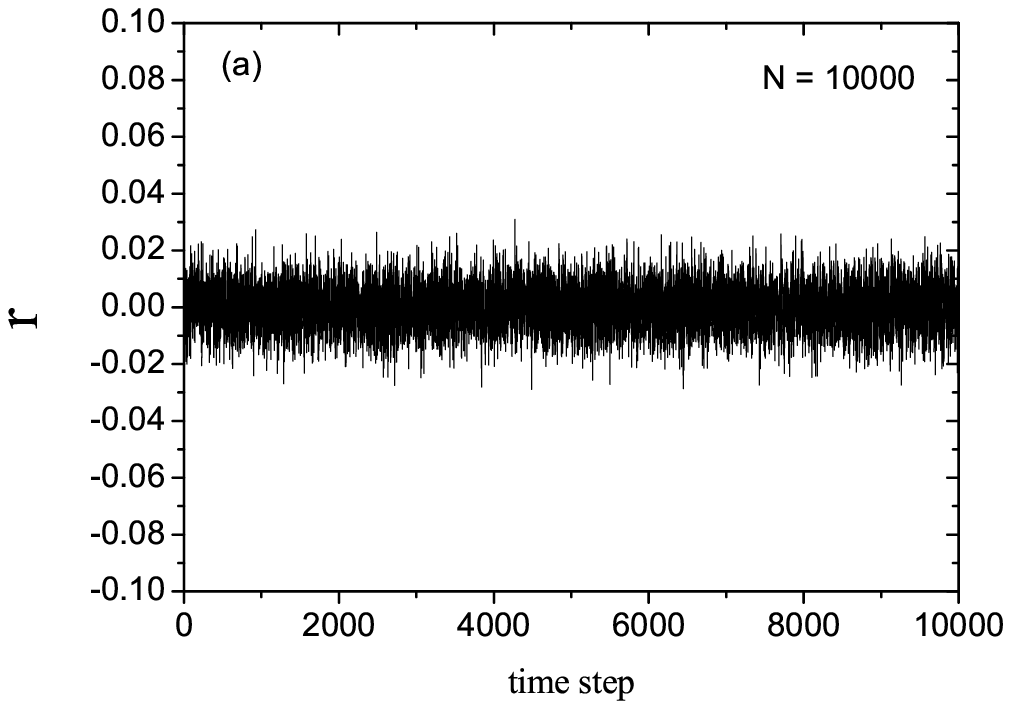}
\includegraphics[width=0.4\textwidth]{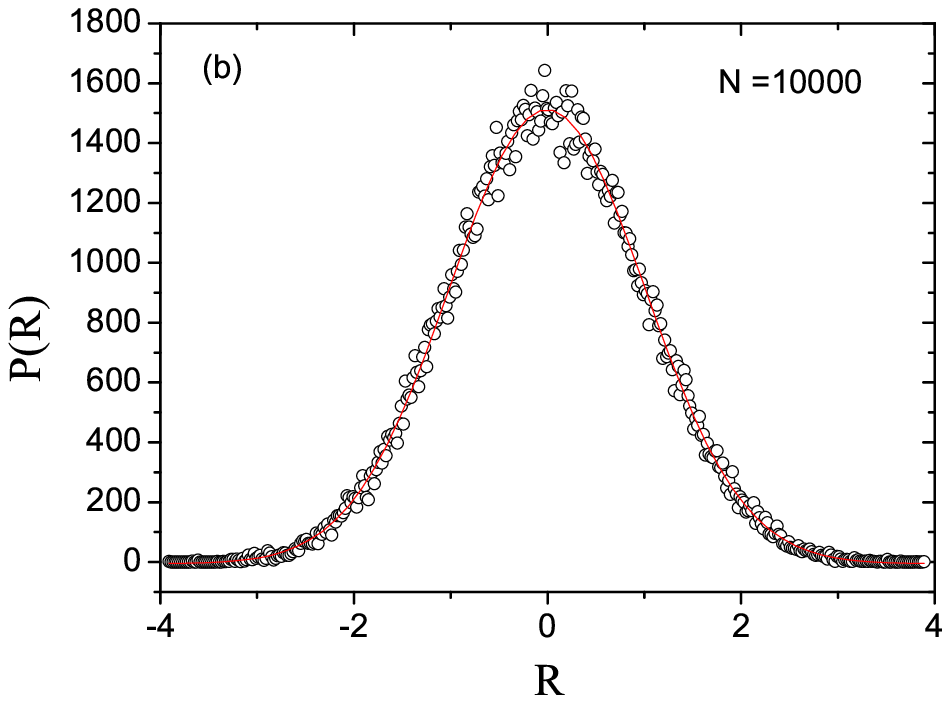}
\includegraphics[width=0.4\textwidth]{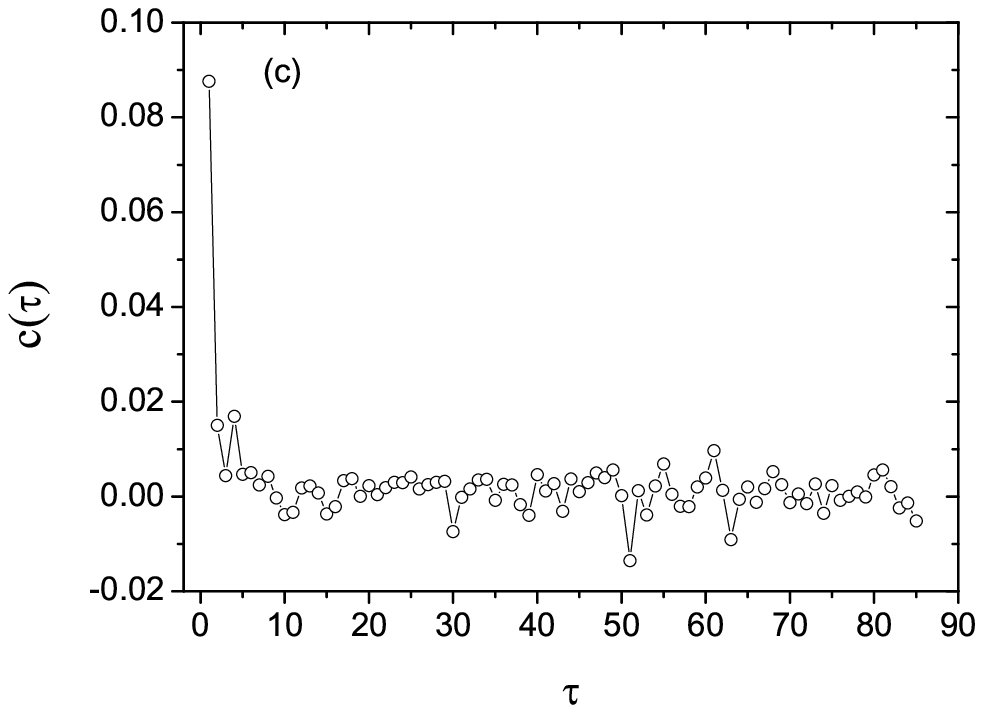}
\caption{(a) Time series of average opinion with the total time steps is $t=10000$, (b) the distribution functions $P(R)$, and (c) the autocorrelation function $c(\tau)$. The parameters used in the simulation are $p(n_{init}=1)=0.95$, $N=10000$, $T=1.0$ and $L=10000$. The parameter $J_{ij}$ and $K_{ij}$ are positive numbers which are not larger than $1$ in whole simulations. All the results in this paper are obtained over $20$ realizations of the social networks.}
\label{result_1}
\end{figure}

At first we investigate the time series of average opinion, as illustrated in Fig.~\ref{result_1}(a). It shows there exists the fluctuation around the average opinion $r = 0$. In order to compare the fluctuation of different scales, the time series have been normalized according to
\begin{equation}
R(t)= \left(r(t)-\left< r(t)\right> _{\tau}\right)/\delta \left( r(t) \right), \nonumber
\end{equation}
where $\left< r(t) \right>_{\tau}$ and $\delta(r(t))$ denote the average and the standard deviation over the period considered, respectively. In Fig.~\ref{result_1}(b), we present the distribution functions $P(R)$ associated with the time series. It is clear that this function $P(R)$ is a Gaussian form.

We calculate the autocorrelation function $c(\tau)$ of our model. For a time series of $L$ samples, $r(t)$ for $t=1,2,\ldots,L$, $c(\tau)$ is defined by
\begin{equation}
c(\tau) = \frac{\sum_{t=1}^{L-\tau}(r(t)-\bar{r})(r(t+\tau)-\bar{r})} {\sum_{t=1}^{L-\tau}(r(t)-\bar{r})^{2}},
\label{eq04}
\end{equation}
where $\tau$ is the time delay and $\bar{r}$ represents the average over the period under consideration. Fig.~\ref{result_1}(c) shows the result of autocorrelation function of our model. It is found that $c(\tau)$ decreases rapidly in very small rang of $\tau$. It means the system has short-time memory effects. As is now well known, the stock market has nontrivial memory effects~\cite{simulation-2}. For example, the autocorrelation funciton of Dow Jones (DJ), also in the small rang of $\tau$, decreases rapidly from $1$ to $0$. From this point, perhaps our model is helpful to understand the financial markets.

\section{The influence of external field\label{sec4}}

\begin{figure}
\includegraphics[width=0.3\textwidth]{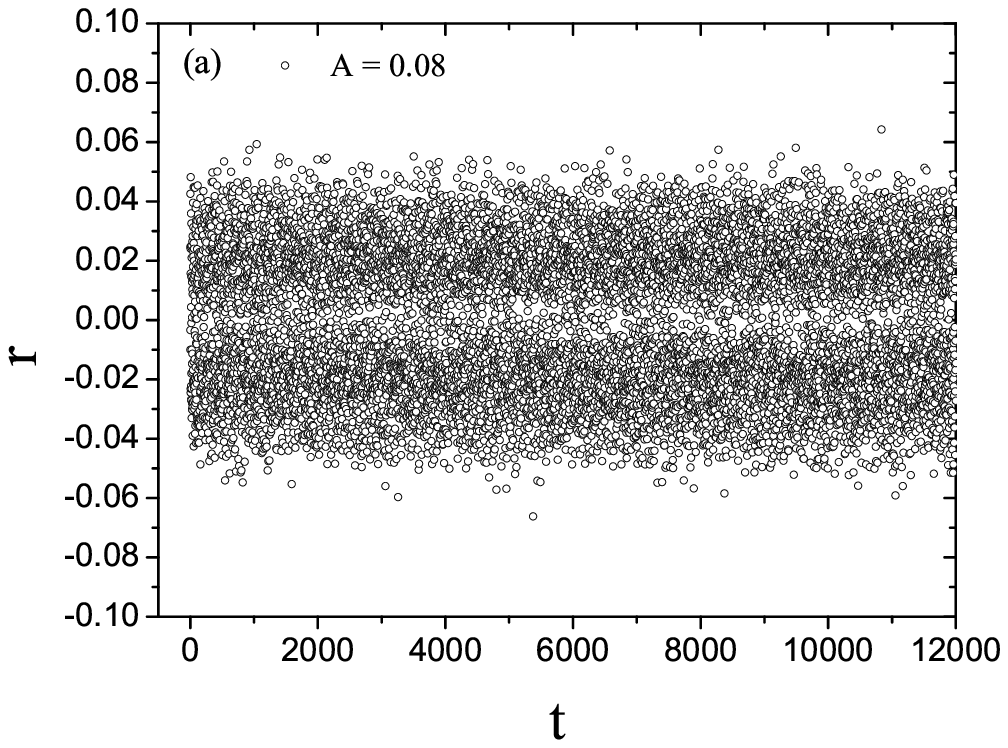}
\includegraphics[width=0.3\textwidth]{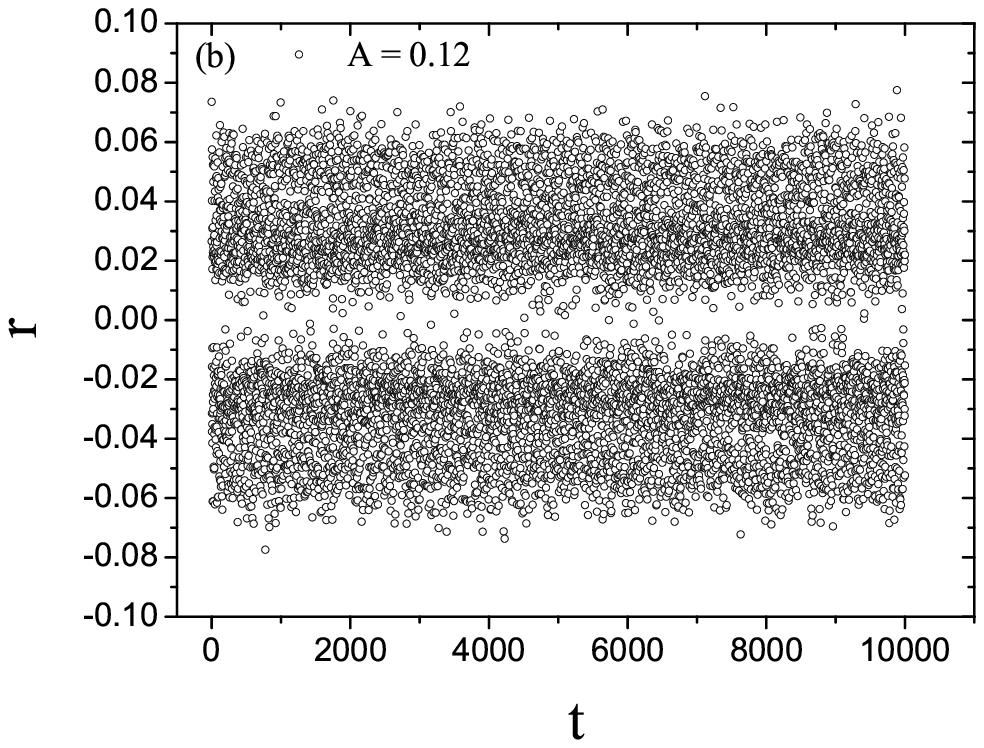}
\includegraphics[width=0.3\textwidth]{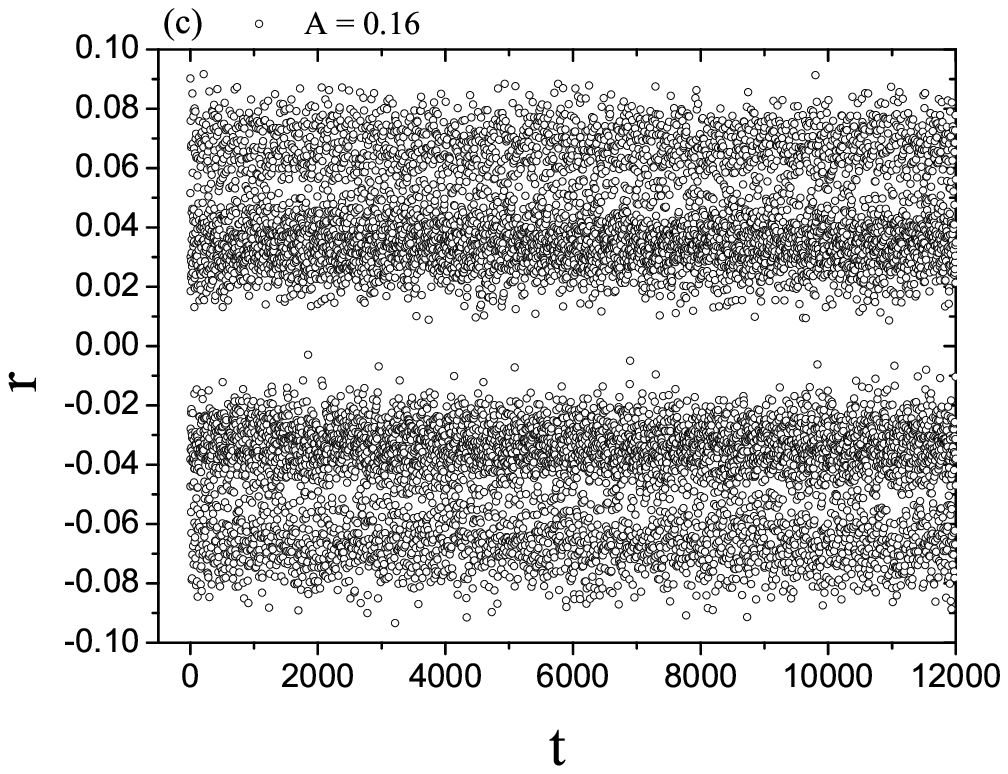}
\includegraphics[width=0.3\textwidth]{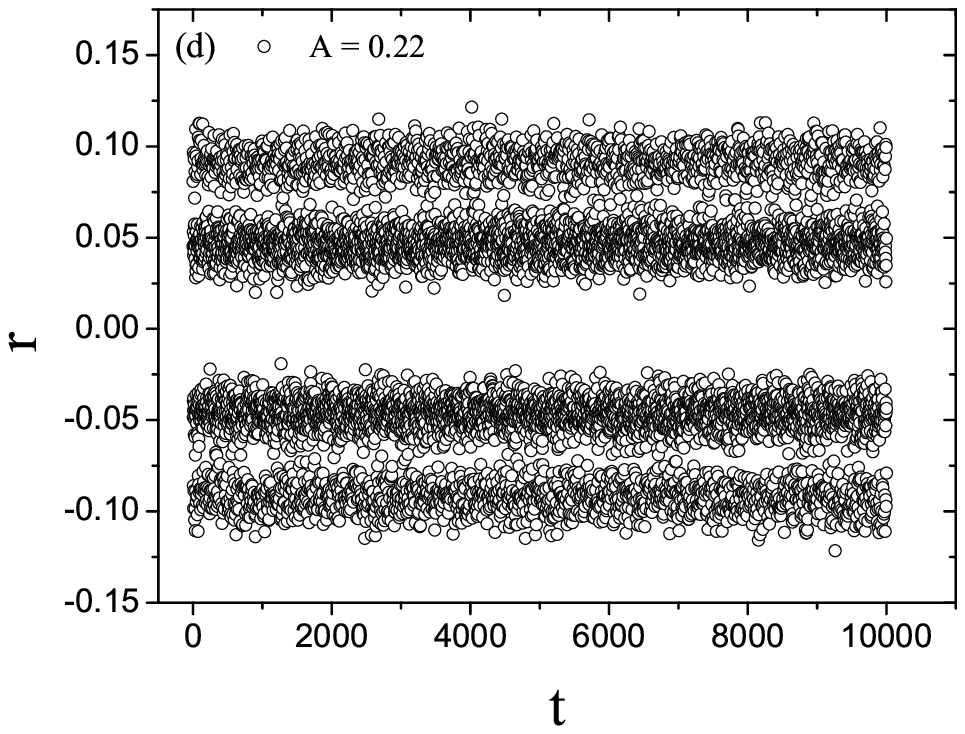}
\includegraphics[width=0.3\textwidth]{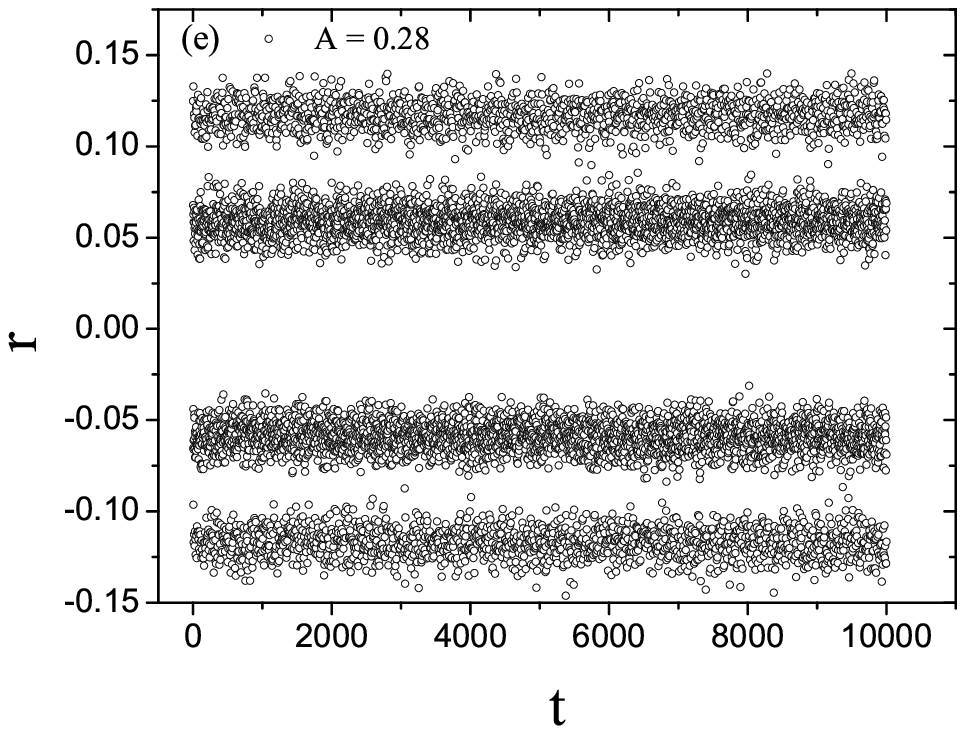}
\includegraphics[width=0.3\textwidth]{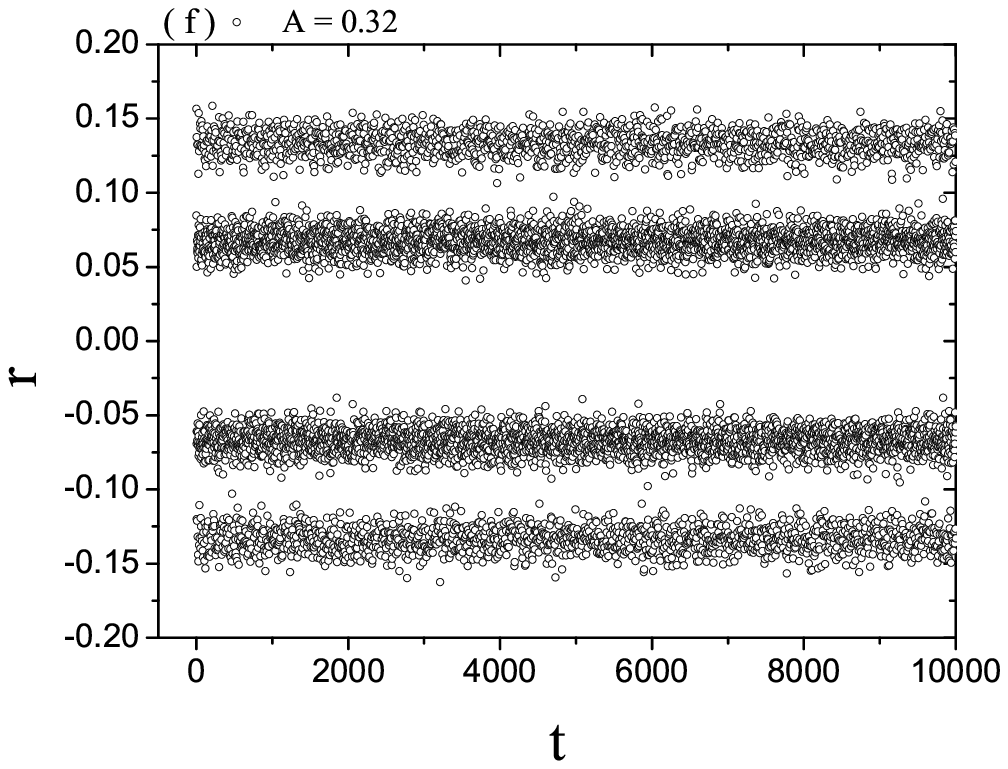}
\caption{Time series of the average opinion with different values of amplitude $A=0.08$, $0.12$, $0.16$, $0.22$, $0.28$, $0.32$. Parameters are $T=1.0$, $\omega=\pi/3$, and $\varphi=0$.}
\label{fig3}
\end{figure}

In order to explore what phenomena maybe happen to system under the influence of external field. We add a period external field to the energy potential $\epsilon_{i}$,
\begin{equation}
\epsilon_{i} \left[ s'|\sigma_{N}(t) \right] = -sh_{i} \left( \sigma_{N}(t) \right) - s^{2} \theta_{i} \left( \sigma_{N}(t) \right) - s \left[ A\cos (\omega t+\varphi) \right],
\label{eq05}
\end{equation}
where $A$ is the amplitude of period external field, $\omega$ is frequency and $\varphi$ denotes the initial phase of external field.

We investigate the effect of amplitude $A$ by fixing other parameters. In Fig.~\ref{fig3} we plot the time series of the average opinion $r(t)$ under different values of $A$. It is obvious that the distribution functions have a remarkable change with increasing $A$. With increasing strength of external field, the average opinion comes into several discrete parts. For small amplitude $A=0.02$, $P(R)$ is still a Gaussian form. When $A=0.08$, it begins to appear two fluctuation around nonzero symmetric values of average opinions. Then, four nonzero average opinions appear at $A=0.16$. Note that the intervals among the discrete average opinions increase with increase in the strength $A$ of external fields. Fig.~\ref{fig3} gives the process from two wave crests to four independent parts. And the average opinion of the whole system will jump from one part to the other parts at all times.

\begin{figure}
\includegraphics[width=0.5\textwidth]{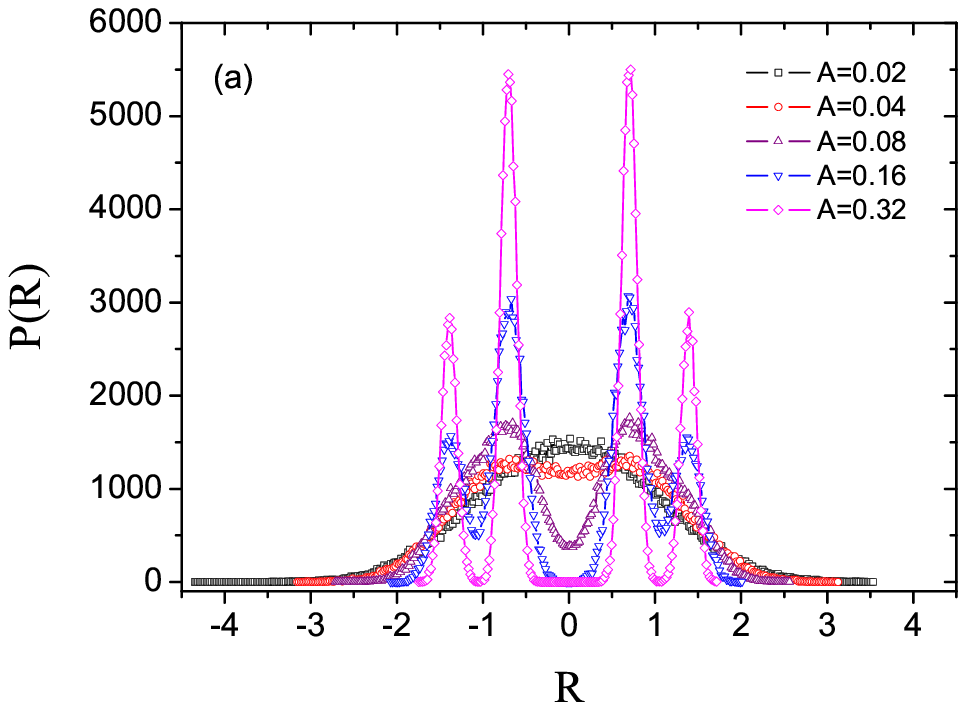}
\includegraphics[width=0.5\textwidth]{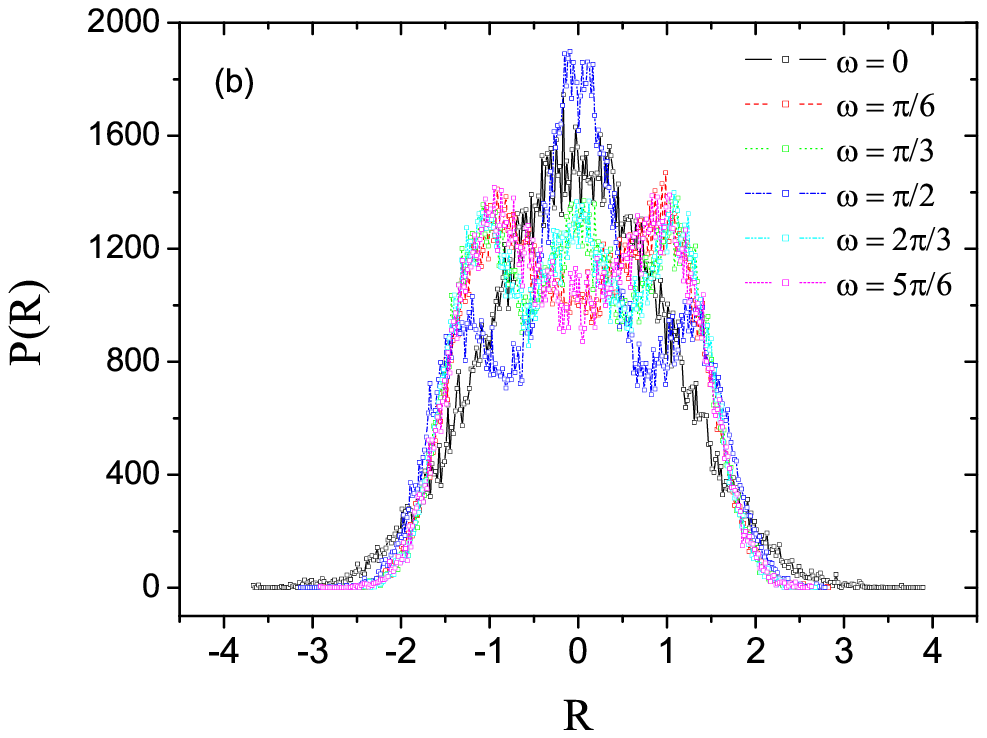}
\caption{(a) The distribution functions $P(R)$ of average opinion time series under different amplitudes $A$. Parameters are $T = 1.0$, $\omega = \pi/3$, and $\varphi = 0$. (b) $P(R)$ for different frequencies $\omega$. Parameters are $A=0.06$, $\varphi=\pi/2$, and $T=1.0$.}
\label{fig4}
\end{figure}

In Fig.~\ref{fig4}, we present the distribution function $P(R)$ of the average opinion. Again, it is easy to verify that the average opinions oscillate among serval separate symmetric nonzero values under the external periodic driving force [see Fig.~\ref{fig4}(a)]. A similar oscillation behavior is observed for simulation on the influence of the frequency $\omega$ which is shown in Fig.~\ref{fig4}(b). Noted that $P(R)$ for the frequency $\omega = \pi/3$ is same to the case for $\omega=2\pi/3$, and the same distribution is observed between $\omega=\pi/6$ and $\omega=5\pi/6$. But there are distinct difference for $\omega=0$ and $\omega=\pi/2$. It indicates a possible period $\pi$ in the case of fixed other parameters.

\begin{figure}
\includegraphics[width=1\textwidth]{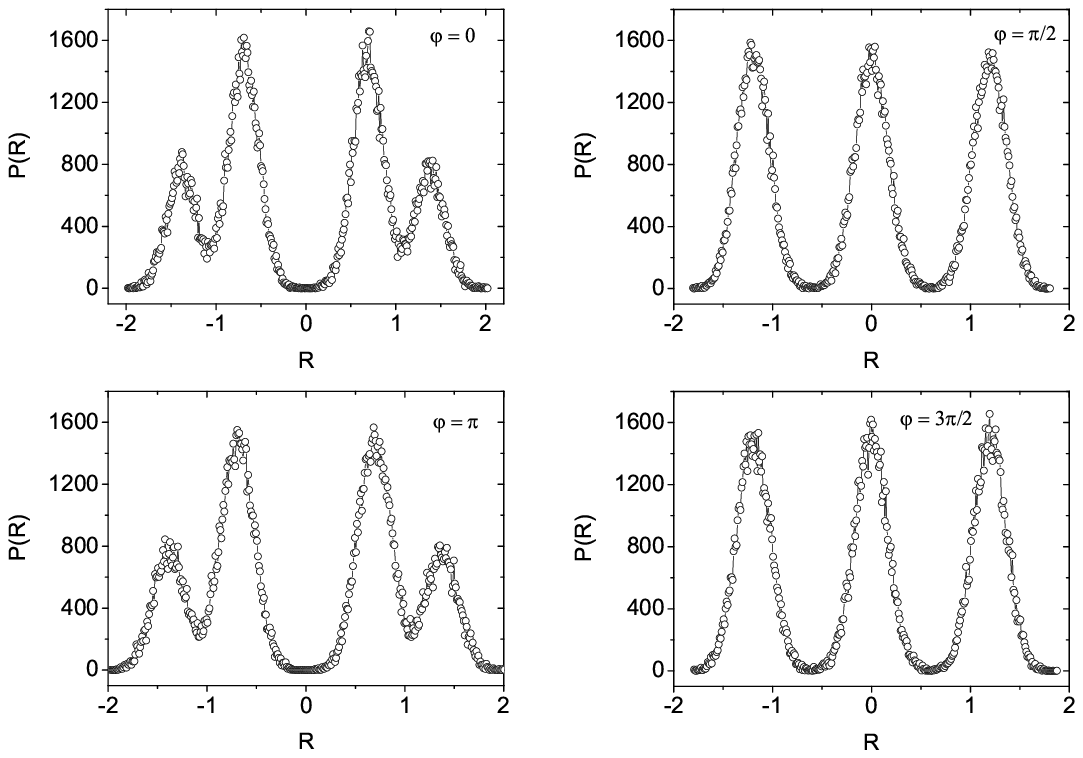}
\caption{The distribution functions $P(R)$ of average opinion time series under different initial phases $\varphi$. Parameters are $A=0.16$, $\omega=\pi/3$, and $T=1.0$.}
\label{fig5}
\end{figure}

Fig.~\ref{fig5} shows the distribution functions $P(R)$ of average opinion time series for different initial phases $\varphi$. For $\varphi =0$, the average opinion vibrates among four symmetric nonzero values. When $\varphi$ increases to $\pi/2$, clearly, the average opinion comes into a $3$-value oscillation. Additionally, note that the distribution functions is almost same for $\varphi=0$ and $\varphi=\pi$ (or $\varphi=\pi/2$ and $\varphi=3\pi/2$). Again, one can conjecture $P(R)$ is a $\pi$-period behavior. We also observe the system's average opinion time series only have two types of distribution functions in different values of initial phases $\varphi$.

\begin{figure}
\includegraphics[width=0.5\textwidth]{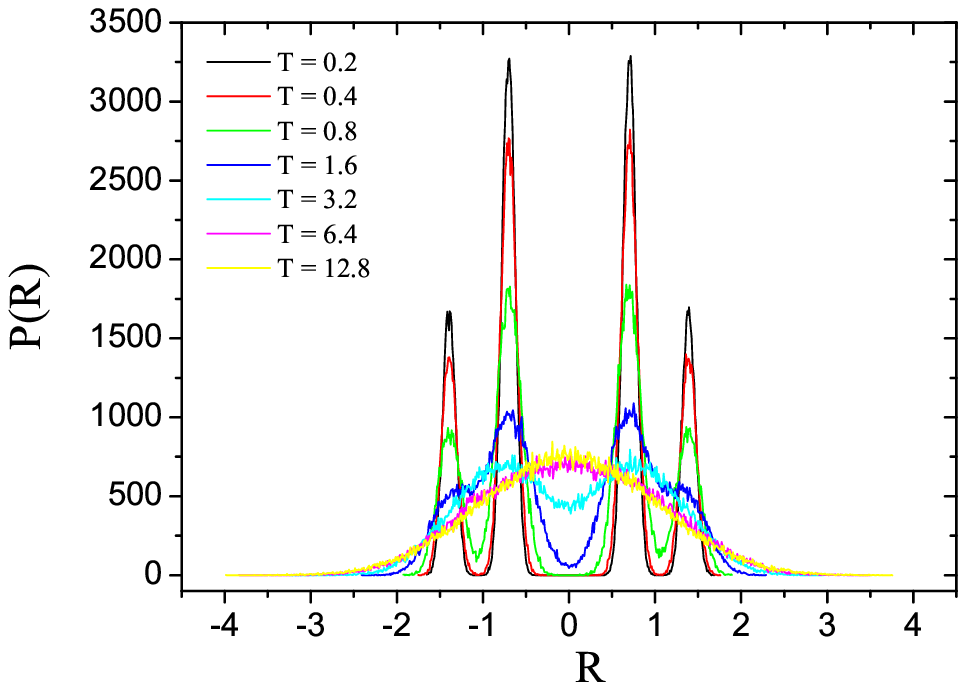}
\caption{The distribution functions $P(R)$ of average opinion time series under different interior thermo-noises $T$. The parameters used in the simulation are $A=0.16$, $\omega=\pi/3$, and $\varphi=0$.}
\label{fig6}
\end{figure}

Another important parameter for the systems is the interior thermo-noise $T$. We explore its effects with (or without) external fields. It is found that there is not remarkable influence on the system without external field. Contrarily, in the case of external field, $P(R)$ shows a similar oscillation with it in Fig.~\ref{fig4}(a) (see Fig.~\ref{fig6}). Note that their influences are opposite. In Fig.~\ref{fig6}, with increasing $T$ the forms of $P(R)$ transform from four-peak to two-peak gradually, and merge into only one-peak at last. At the same time, the average opinion $r$ is expanded from some separate regions to the whole more expansive scale for larger $T$.

By comparing the Fig.~\ref{fig4}(a) with the Fig.~\ref{fig6}, it is clear that the amplitude $A$ and interior thermo-noise $T$ have opposite effects acting on the systems. It looks like a couple of contradictory parameters, even though both lead to the split phenomena of the distribution of average opinion $P(R)$ and the nonzero average $R$.

It exists similar behaviors in the Ising ferromagnetic systems. In Ising model, the order-disorder transition is a second order transition. It will be a non-zero magnetization $\pm|M_{sp}|$ for a finite system. There is a nonzero probability for ever that the system from near $+|M_{sp}|$ to near $-|M_{sp}|$, and vice versa~\cite{external-1}. In our model under the influence of external field, it is also observed the phenomena of phase transition caused by $T$ (or by $A$), which is similar to the Ising paramagnetic-antiferromagnetic transition.

As discussed above, the energy potential increases with increasing $T$, and the system's entropy becomes larger (more disordered). But the external field tends to restrict the disordered effects in the system and reduces the disordered strength into several separate regions.

\section{Conclusion\label{sec5}}

In the present work we introduce Blume-Emery-Griffiths model on opinion formation with three-state. Considering the characters of real social systems, we construct a social network to link between agents. In this BEG model, each person's opinion is influenced not only by his specific local information from his neighbors but also by the average opinion of the whole network.

Moreover, we focus on the behaviors of BEG systems under external perturbation. The simulation results show that this system is sensitive to the external field. As discussed in Sec.~\ref{sec3}, the parameters in the external periodic perturbation, such as amplitude $A$, initial phase $\varphi$, and frequency $\omega$, have obvious impacts on the opinion systems. Besides, the effect of the amplitude $A$ or interior thermo-noise $T$ is similar to the Ising paramagnetic-antiferromagnetic transition, and the influence acted on systems from $A$ and $T$ is opposite.

\end{document}